# Architectural control of freeze-cast ceramics through additives and templating

Etienne Munch[2], Eduardo Saiz[2], Antoni P. Tomsia[2], Sylvain Deville[1*]

[2] Materials Sciences Division, Lawrence Berkeley National Laboratory, Berkeley, California, USA

[1] Laboratoire de Synthèse et Fonctionnalisation des Céramiques

UMR3080 CNRS/Saint-Gobain CREE, 550, Avenue Alphonse Jauffret, BP 224, 84306 Cavaillon Cedex, FRANCE

* Corresponding author. Email: sylvain.deville@saint-gobain.com

**Abstract**

The freezing of concentrated colloidal suspensions is a complex physical process involving a large number of parameters. These parameters provide unique tools to manipulate the architecture of freeze-cast materials at multiple length scales in a single processing step. However, we are still far from developing predictive models to describe the growth of ice crystals in concentrated particle slurries. In order to exert reliable control over the microstructural formation of freeze-cast materials, it is necessary to reach a deeper understanding of the basic relationships between the experimental conditions and the microstructure of the growing solid. In this work, we explore the role of several processing variables (e.g., composition of the suspension, freezing rate, and patterning of the freezing surface) that could affect the formulation strategies for the architectural manipulation of freeze-cast materials. We also demonstrate, using freeze-cast lamellar structures, how reducing the lamellar thickness by less than half increases compressive strength by more than one order of magnitude.

# I. Introduction

Emerging technologies in energy storage, catalysis or tissue engineering demand new materials with highly specialized and often seemingly incompatible functional properties such as high surface area, low density, and highly specific strength or toughness. Today, a number of researchers are working on the development of new processing routes capable of delivering materials to fulfill these complex requirements. Organized hierarchical structures, often inspired by natural structures and biological materials[1], are rapidly becoming attractive solutions to these challenges. Nonetheless, the complex architecture of natural materials continues to be extremely difficult to reproduce in a controlled manner. As a consequence the vast majority of current processing techniques yield materials with functional properties that are disappointing when compared to those of natural biomaterials.

Freeze-casting is a versatile processing technique that solidifies a colloidal ceramic suspension under conditions in which the freezing dispersant interacts with suspended ceramic particles, pushing them away from the advancing solid front.[2-5] After the frozen liquid has been removed by sublimation, a replica of the solidified solvent is formed. Our group has recently demonstrated how freeze-casting can be used to fabricate porous lamellar ceramics with complex structures which resemble the inorganic component of nacre.[6] The freezing of concentrated colloidal suspensions is a complex physical process involving a large number of parameters[7-10]. These parameters provide unique tools to manipulate the architecture of freeze-cast materials at multiple length scales from the nano to the macro levels in a single processing step. However, in order to exert reliable microstructural control, it is necessary to obtain a deeper understanding of the basic relationships between experimental conditions (e.g., freezing rates, composition of the dispersion) and the microstructure of the growing solid. In this work, we explore the role of several processing variables (e.g., composition of the suspension, freezing rate, and patterning of the freezing surface) that could be used in the formulation of effective strategies for multiscale manipulation of freeze-cast materials.

While we are not attempting to fully quantify the precise influence of each given parameter, we have illustrated how they can provide a new toolbox for fabricating novel freeze-cast structures. As an example, we will show how reducing the thickness of freeze-cast lamellae by less than half results in increased compressive strength by more than one order of magnitude. As demonstrated by our experiments, our technique not only promises to guide the design and fabrication of new hierarchical and porous materials, but it can also benefit any scientific field of research, including physics, chemistry, biology, or geology, where the solidification of concentrated colloidal suspensions plays a critical role.

## II. Experimental procedure

Slurries were prepared by mixing distilled water with a small amount of ammonium polymethacrylate anionic dispersant (1 wt% of powder) (Darvan C, R. T. Vanderbilt Co., Norwalk, CT), an organic binder (polyvinyl-alcohol), alumina powder (Ceralox SPA05, Ceralox Div., Condea Vista Co., Tucson, USA), and different types of additives in various proportions (see the results and discussion section for the precise amounts). The following additives have been used: sucrose, minimum 99.5% (Sigma, St. Louis, MO, USA); trehalose, D(+)-trehalose dihydrate from corn starch (Sigma, St. Louis, MO, USA); glycerol (Mallinckrodt, Inc., St Louis, MO, USA); ethanol, ethanol reagent denatured HPLC grade (Sigma, St. Louis, MO, USA); sodium chloride, Morton iodized salt (Morton, Philadelphia, PA, USA); gelatine, Gelatine Knox Original, unflavored (*Knox*, Kraft Food, USA); citric acid, monohydrate 99.5+% (*Alfa Aesar*, Ward Hill, MA, USA).

Slurries were ball milled for 20 hours with alumina balls and de-aired by stirring in a vacuum desiccator until air bubbles were completely removed (typically 30 minutes). The powder used in the study has a specific area of 8.1 m$^2$/g and an average grain size of 400 nm (data provided by manufacturer).

Slurries were frozen by pouring them into a Teflon® mold (18 mm diameter, 30 mm length) placed at the top of a cold finger (copper rod), and cooled by liquid nitrogen (Fig. 1). Freezing kinetics were controlled by a heater placed on the metallic rods, and a thermocouple placed on the side of the mold. An additional vibration was applied to homogenize the cooling suspension. Samples were freeze-dried (Freeze Dryer 8, Labconco, Kansas City, MI) for 24 hours. The green bodies thus produced were sintered in air for 2 hours at 1500°C, with heating and cooling rates of 5°C/minute (1216BL, CM Furnaces Inc., Bloomfield, NJ). The microstructure of the samples was analyzed by optical and environmental scanning electron microscopy (ESEM, S-4300SE/N, Hitachi, Pleasanton, CA). The wavelength was measured in parts of the sample where a constant thickness of the ceramic lamellae was established.

Compression tests have been performed on cubes (4×4×4 mm$^3$) on an MTS 810 material test system, at a constant displacement rate of 10 μm/s$^{-1}$. Five specimens were used for each experimental condition.

The samples processed in this study have dimensions typically in the centimeter range (Fig. 2). The planar dimensions are only limited by the experimental setup, and no scaling up issues should be expected. The achievable thickness is limited by the ability to maintain a homogeneous structure along the thickness. An improved experimental setup [11] allowed 40 mm thick freeze-cast samples to be processed, exhibiting a homogeneous structure along the thickness.

### III. Results and discussion

### III.1 Control of lamellae orientation

Among the common difficulties encountered in the pursuit of hierarchical materials is the control of the number of dimensions in which the structures are ordered. Typical freeze-cast materials exhibit only a one-dimensional ordering of the structure: the porosity is continuous along the propagation direction of the solid-liquid interface, but the pores are randomly oriented in a plane perpendicular to the ice-front propagation. A second dimension can be added for polymeric materials by performing several freezing steps[3], although this approach cannot be used with particle suspensions. Obtaining such an improved control in a single-step requires a different strategy. The initial instance of directional freezing is usually characterized by the formation of a planar ice front, which nucleates and grows on the cold finger[9]. This front moves very fast, trapping the ceramic particles. Subsequently, a planar-to-cellular-to-lamellar transition sequence occurs, and a steady state is achieved in which lamellar ice grows while expelling the ceramic particles that accumulated in the layers between ice crystals[9]. When a well-polished cold surface is used, the planar-to-lamellar transition results in the formation of randomly oriented lamellar domains of sizes on the order of 200-500 µm (Fig. 3a). This is very similar to what happens when fast nucleation is induced by pouring the suspension of a surface at temperatures far below the freezing point of the suspension. Using simple, unidirectional patterns on the cooling surface, it is possible to manipulate the direction of the ice lamellae arising from the planar to lamellar transition and to obtain long-range ordering with well-oriented structures on the length scale of the whole sample (Fig. 3b), suggesting the ordering is controlled by nucleation.

The lamellae orient perpendicularly to the linear patterns, introducing a second dimension to the ordering of the structure. The crystals' perpendicular orientation to the pattern may be explained by a preferred nucleation at sharp-angles of the pattern. Further investigations will be necessary to clarify these aspects.

Patterns require the right dimension to achieve lamellae alignment. The thickness of the planar ice layer, the particle size, and the speed of the ice front determine the wavelength of the lamellar structure. Pattern sizes vary between 40 to 100 µm for 10 to 50 µm lamellae, respectively.

More complex lamellae orientations that can be dictated by functional requirements are difficult to control by simple cold-finger patterning. We have developed a simple epitaxy technique that places a mold filled with a pure solvent (water, in our case) on a cold finger. A second mold filled with the freeze-casting suspension is subsequently placed on top. When the cold-finger temperature reaches the solidification point, pure ice crystals grow in the patterned mold and extend onto the ceramic suspension, resulting in a macroscopic ice pattern directed by the mold surface features. This way, it was possible to replicate (Fig. 3c) the circular macrostructure of natural lamellar materials such as the osteons in bone[12] or the Euplectella sponge spicules[13], adding another level of structural control at macroscopic dimensions that can promote additional mechanisms to manipulate the mechanical response.

### III.2 Using additives to control the structure

Controlling the structure at the micron and submicron levels requires the manipulation of the thickness and roughness of the growing ice crystals. Again, comparison with natural materials points out the importance of several microstructural features: the thickness and separations between the ceramic layers (the structural wavelength), the roughness of the ceramic walls at the microscopic and submicroscopic levels, and the presence of inorganic ceramic bridges between lamellae[14,15]. These features play a crucial role in the mechanical response of natural layered materials such as nacre. A processing technique able to provide structural control from the macro- to the nano-levels will make it possible to fabricate truly biomimetic structures.

It has already been shown how the solid content of the solution and the ice-front speed can be manipulated to control the wavelength of the lamellar structures resulting from freeze-casting[9]. The strategy explored in this work is the use of chemical additives that will affect the growth kinetics and microstructure of ice as well as the topology of the ice-water interface. Additives influence several physical parameters controlling the microstructure of the growing ice crystals; for example, additives can change the phase diagram of the solvent, the value and anisotropy of the solid/liquid and particle/liquid interfacial energies, the degree of under-cooling ahead of the liquid front, the viscosity of the solvent, and the forces between ceramic particles in the suspension. Some of these effects are interrelated, but as a result, the morphology of the ice crystals and their roughness can be manipulated down to submicron levels.

For our work, we selected additives commonly used in fields as diverse as cryopreservation, ice cream processing, and marine glaciology. Some additives are well-known antifreezers, including sodium chloride, carbohydrates such as sucrose and trehalose, glycerol, and ethanol. Others such as gelatin were used to affect the viscosity of the suspension. Finally, we reduced pH by adding citric acid to manipulate inter-particle forces[16].

Several of the additives used in this work (e.g., NaCl) have a eutectic phase diagram with water. As a consequence, the solid and liquid regions are separated by a zone in which solid and liquid can coexist (Fig. 4) during ice growth in a temperature gradient. The height of this zone depends on the temperature gradient and on the particular phase diagram (the slope of the liquidus line). As the freezing front advances, a second phase will precipitate. Like the ice, this is a fugitive phase that will disappear during drying or sintering, but its morphology will have a strong influence on the microstructure of the porous scaffold. Through their influence on the interfacial energies and the solidification path, additives will dictate whether the materials' final architecture can be switched (Fig. 5a-d) from lamellar (e.g. no additive, trehalose, or sucrose) to cellular (e.g. gelatin, glycerol, or combining sucrose with a reduction of the pH using citric acid), or to a lamellar structure with a bimodal pore width distribution (ethanol), where pores with a very large cross section (a few hundred microns) are observed in addition to the basic lamellar structure previously shown. Such a structure exhibits interesting similarities with natural structures such as wood[17,18], which provides an additional degree of hierarchy to the structure, and could prove to be useful in applications (catalysis, filtration) where a compromise should be found between specific surface area and the accessibility of fluids through the pore channels.

Besides the overall architecture, additives can also be used to modify the roughness of the ceramic walls (Fig. 5e-h). For example, the water-NaCl diagram (Fig. 4) has a eutectic point relatively close to the melting point of ice (-21°C). During cooling, crystalline NaCl·2H$_2$O will precipitate. This second phase will also expel the ceramic particles during solidification and subsequently disappear during firing at high temperature, resulting in a sharp, faceted microstructure (Fig. 5-g). On the other hand, sucrose and trehalose water diagrams [19,20] are very similar, and for the structures used in this work, result in a second glassy phase when the temperature reaches -40 to -50°C. Correspondingly, samples prepared with these two additives exhibited rounded, smooth features. In addition, the additives modified the interfacial energies and the degree of supercooling ahead of the freezing front, affecting the overall morphology of the growing ice crystals. Sucrose and trehalose promote the evenly distributed growth of homogeneous and regular dendrites over the surface of the lamellae (Fig. 5-h). The relative dimensions (width and height) of the dendrites can be modified by increasing the concentration of an additive in the suspension; for example, the dendrites become thinner at higher concentrations. In contrast, ethanol leads to very different topography, with pocket-like dendrites covering the surface of the lamellae (Fig. 5-f). The variations in surface roughness are also reflected in the specific surface area of the material. Additions of sucrose, ethanol, and trehalose lead to different values of specific surface areas according to their different roughnesses (sucrose 0.47 m$^2$.g$^{-1}$, ethanol 0.36 m$^2$.g$^{-1}$, and trehalose 0.41 m$^2$.g$^{-1}$).

It must be pointed out that, despite the obvious microstructural differences, the relationship between the structural wavelength and the average speed of the ice front seems almost independent of the additive used (Fig. 6). The only exception is ethanol; the underlying reasons for this behavior are not yet understood, but it could possibly be related to the coexistence of a liquid and a solid (crystal) phase at relatively low temperatures (e.g., -72°C) and the complexity of crystal growth from the ethanol-water solution.

**III.3 Using additives to control lamellar bridges**

Another critical microstructural feature is the presence of ceramic bridges between adjacent lamellae. The bridges play a crucial role in the mechanical response, preventing Euler buckling during compression, and hindering crack propagation in the direction parallel to the lamellae. The velocity of the ice front during the freeze-casting process is below the critical velocity for particle trapping[9]. During the freezing of highly concentrated suspensions, the splitting and healing ice crystal tips can result in the encapsulation of ceramic particles and the formation of bridges between lamellae. Each additive effects interfacial tension and interparticle forces leading to a particular bridge structure. For example, trehalose promotes a high density ($<\rho> = 27.3$ cm$^{-1}$) of thin bridges ($<w>=1.6$ μm), whereas sucrose leads to a low density ($<\rho> = 16.1$ cm$^{-1}$) of wider bridges ($<w>=2.2$ μm).

### III.4 Structure/properties relationships

By controlling the solid content, the cooling rate and the use of diverse additives, freeze-casting of water-based suspensions allows the design of inorganic scaffolds with a wide range of structures ranging from lamellar to cellular, and from low- to high-density bridging between lamellae, and having various levels of roughness at a local scale (Fig. 5). To illustrate the importance of the materials' structure, the comparison between the compressive strengths of a highly uniform lamellar alumina prepared by adding 4 wt % sucrose in water and a cellular structure formed using 10 wt % sucrose in water under acid conditions (pH=2.5) is shown in Fig. 7. Both structures have the same relative porosity and average size of the ceramic wall. However, the cellular structure exhibits a bimodal distribution of wall thickness (Fig. 7a) while the lamellar structure has a very narrow distribution (Fig. 7b). The microscopic roughness of the lamellae is not expected to influence the mechanical behavior of the highly porous materials. As expected, the cellular structure has a relatively more isotropic response, and for the same average lamellae thickness is stronger (Fig. 7c) . However, the thickness of the lamellae has a tremendous impact on the strength of the lamellar material, and decreasing the thickness by less than half increases the strength by almost an order of magnitude (Fig. 7d). This increase may not be fully surprising if we consider the refinement of the structure, the increasing presence of bridges for refined lamellae and the smaller defect size in thinner layers. A systematic assessment of the failure modes in these complex lamellar structures can guide the design of optimized microstrucures.

**III.5 Multiscale control of the structure**

Processing techniques for cellular ceramics usually have little control over a material's characteristics, yielding very specific porous microstructures of controlled pore size and shape. Hence, a processing route must be chosen based on the desired structure. Freeze-casting is a very versatile technique, allowing streamlined control of several important structural features, in a wide range of length scales, each of which can be tailored with some degree of freedom, by different mechanisms. The different features of a freeze-cast ceramic structure and their corresponding controls are summarized in Table 1. Some of these controls, such as the influence of the solidification interface velocity on the lamellae's thickness, or the use of solvents other than water, have been demonstrated in previous papers. Additional strategies to control the structure have been developed with the freeze-casting of polymers, such as consecutive solidification stages using different materials to process biaxially aligned structures[3]; however, it is not always possible to transfer these methods to ceramic freeze-casting.

We have introduced two new powerful strategies: templating the cooling surface and the use of common additives. The templating is completely independent of the other strategies to control the structure of the materials. The use of additives offers a number of advantages over modifying a solvent; for example, additives broaden the number of possible pore-shape morphologies, and the same equipment can be used to process different materials. However, much more work is needed to assess the precise influence of additives, and to determine the degree of a control's interdependence or independence.

**IV. Conclusions**

Processing techniques for porous materials in general and cellular ceramics in particular yield very specific porous microstructures with controlled pore size and shape. However, the manipulation of pore orientation, morphology, and surface topography is typically very limited. Freeze-casting appears to be a very versatile technique, allowing the use of different mechanisms to control several important structural features, in a wide range of scales, all of which can be tailored independently. The various physicochemical processes controlling the solidification of the colloidal suspensions provide a set of tools to control various features of the structure in a single processing step. This versatile, streamlined process allows the development of truly biomimetic, hierarchical structures. The right combination of experimental freezing conditions and different suspension formulations makes it possible to adjust the orientation of lamellae, and also provides a new means to manipulate the order, the size and shape of pores, the wavelength of the structure, the roughness of lamellae surface, and the density and thickness of bridges between lamellae. This also makes it possible to refine the porous structure, and to consequently manipulate a material's mechanical response.

**Acknowledgments**

This work was supported by the Director, Office of Science, Office of Basic Energy Sciences, Materials Sciences and Engineering Division, of the U.S. Department of Energy under Contract No. DE-AC02-05CH11231

**Figures**

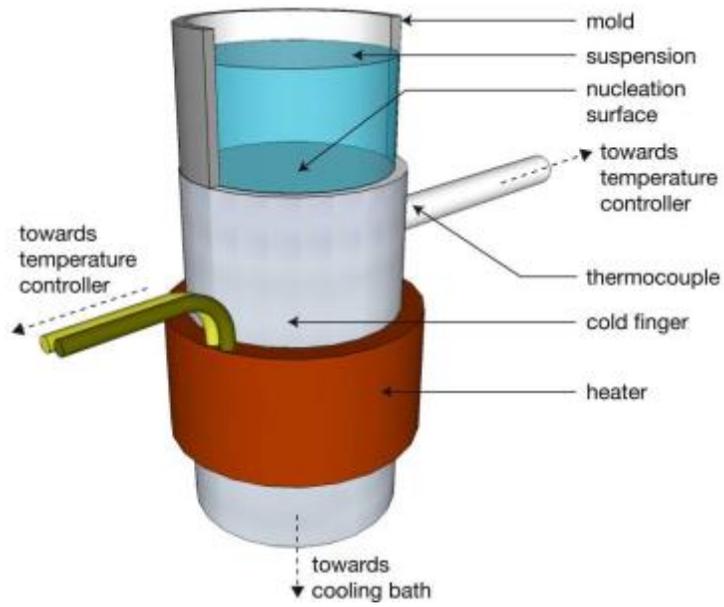

*Figure 1: Experimental setup*

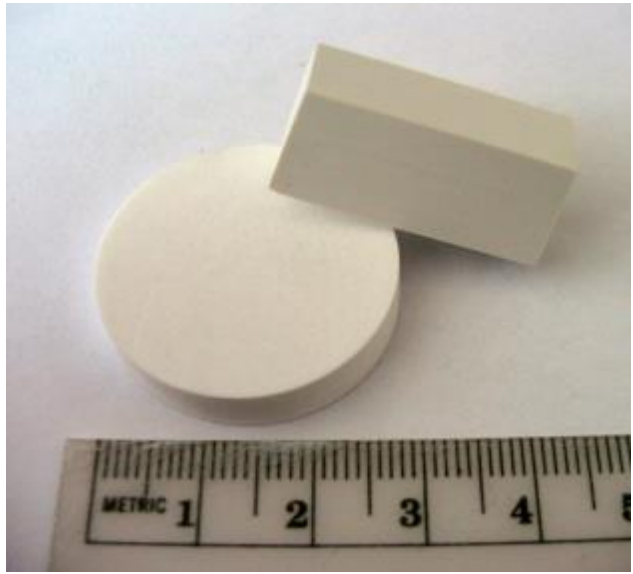

*Figure 2: Typical sample sizes and geometries*

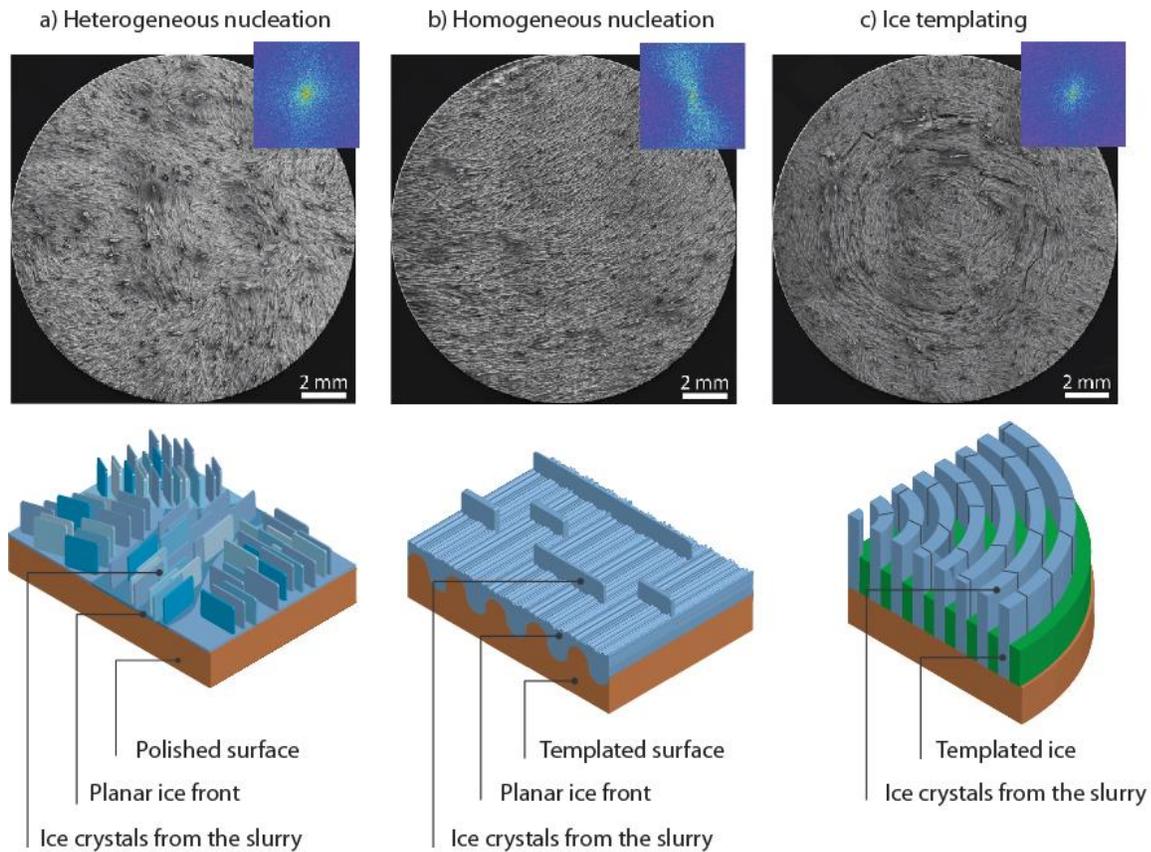

*Figure 3: Control of the structure at a macroscopic level. (a) By using a cold finger with a polished surface, the lamellae organized into random domains several hundred microns in size. (b) When a unidirectional pattern is applied to the cold finger, the lamellae align over macroscopic dimensions (cm). (c) Using a mold filled with water in the cold finger, it is possible to impose complex (e.g., circular) patterns onto the growing ice crystals, adding new dimensions to the order. The insets in the scanning electron micrographs show the Fourier transform of the images and illustrate the lamellae alignment in (b).*

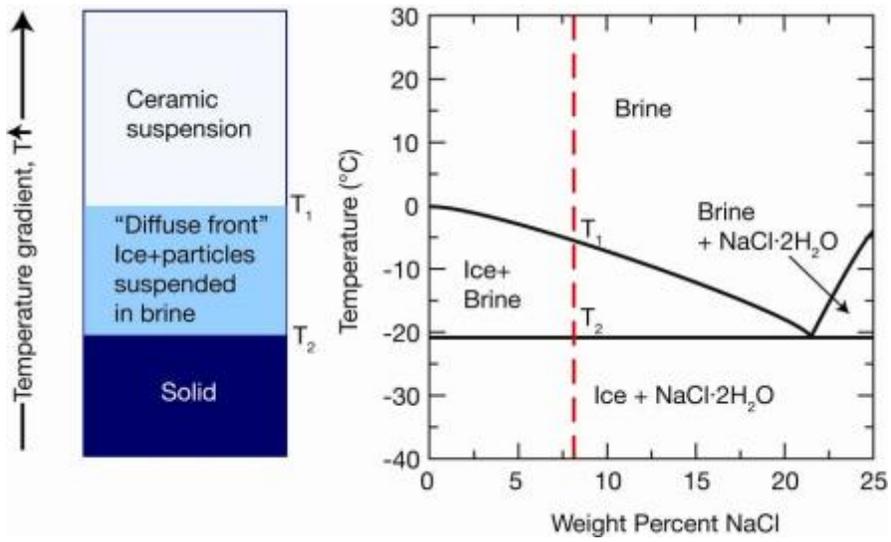

*Figure 4: Water-NaCl phase diagram and corresponding solidification behavior*

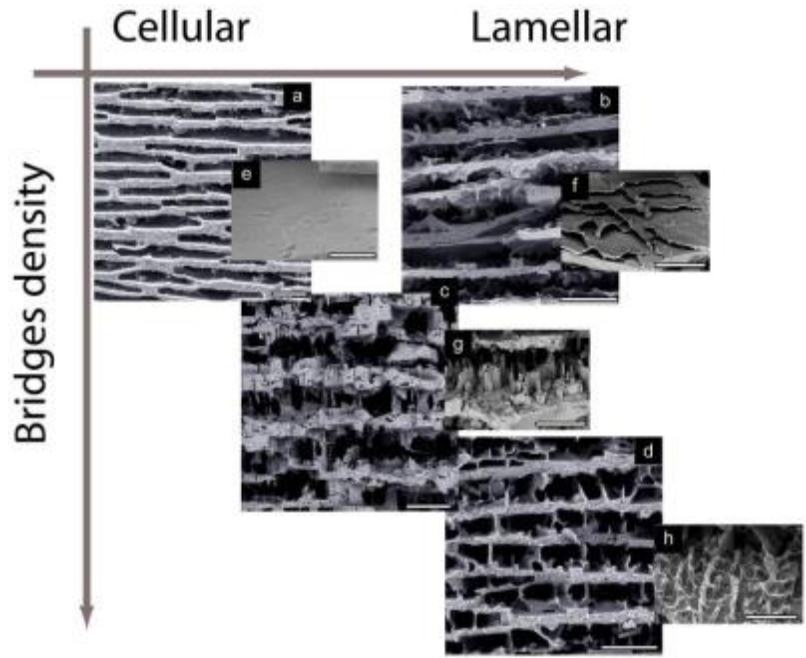

*Figure 5: Effect of additives on a material's architecture. The graph shows how different additives can change the structure from lamellar to cellular, and that the roughness and density of bridges between lamellae can be manipulated. Scanning electron micrographs of: (a) 10 wt.% sucrose in citric water pH=2.5; the structure is cellular with very smooth ceramic walls. (b) 4 wt.% ethanol results in lamellar structure with smooth roughness. (c) 4 wt.% sodium chloride gives a sharp faceted lamellae surface. (d) 4 wt.% sucrose promotes a lamellar structure with microscopic roughness. Insets (e) to (h): Details of the surface roughness of the lamellae. Scale bars: (a) 50 μm (b) 100 μm (c) 100 μm (d) 50 μm (e) 50 μm (f) 100 μm (g) 100 μm (h) 100 μm*

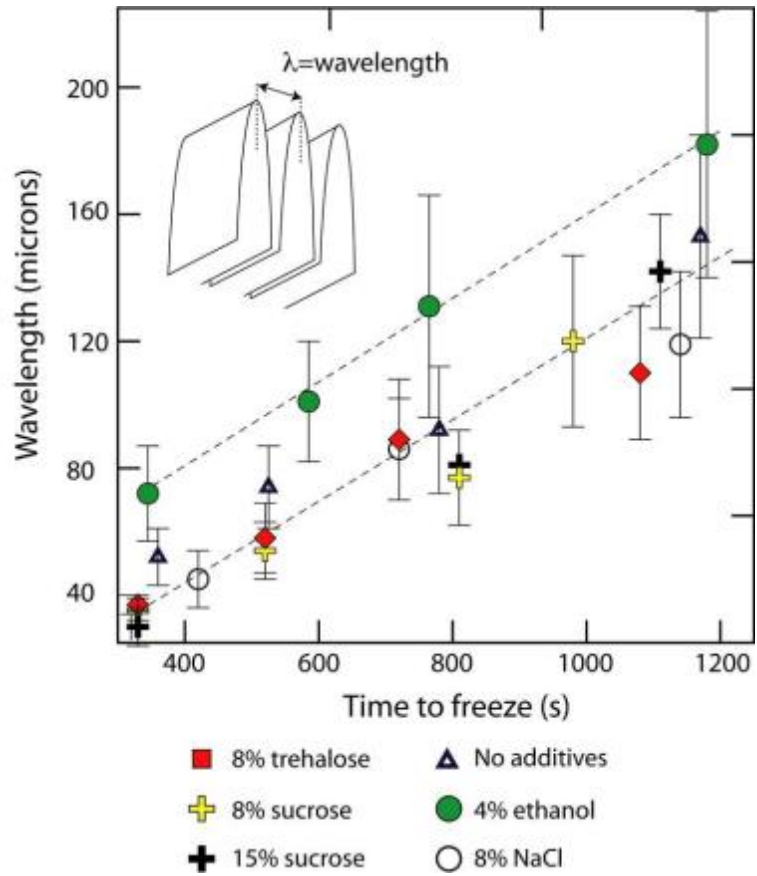

*Figure 6: Wavelength vs. time to freeze for 2 cm tall samples. Error bars represent the standard deviation (>100 measurements for each data point).*

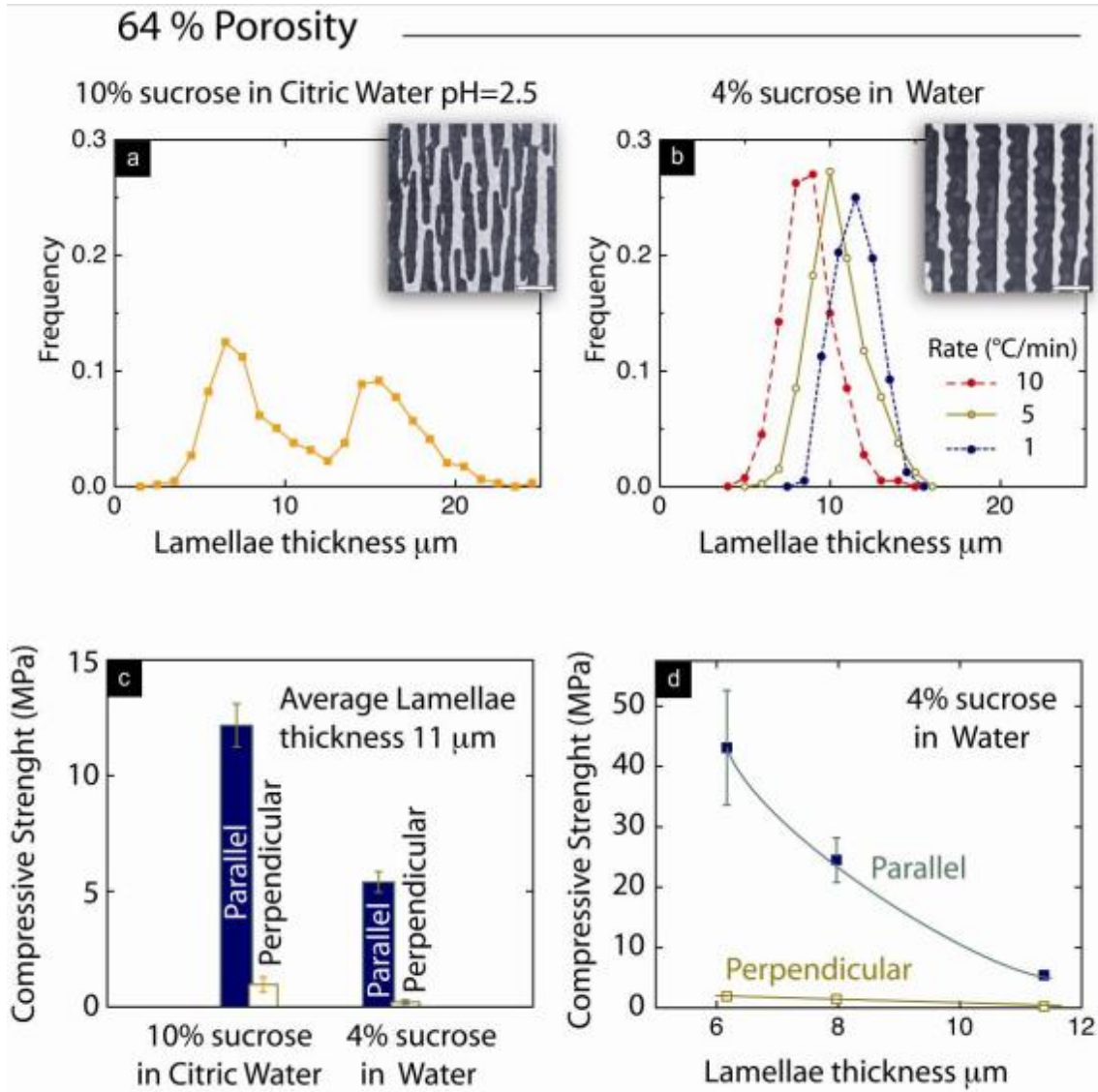

*Figure 7: Link between structure and mechanical properties. (a-b) Normalized distribution of lamellae thickness. Both samples have the same average lamellae thickness, but the sample with 10 wt.% of sucrose in citric water (a) has a cellular structure with a bimodal wall thickness distribution, and the sample with 4 wt.% sucrose in water (b) has a lamellar structure with uniform lamellae thickness that depends on the cooling rate of the cold finger (the faster the cooling rate, the faster the speed of the ice front). The compressive strengths in the directions parallel and perpendicular to the lamellae orientation depend on the structure and the average lamellae thickness. For samples with a more cellular structure, the thickness strength is less anisotropic (c). In lamellar samples, the strength is very dependent on the thickness of the ceramic layers. The*

*compressive strength in the direction parallel to the lamellae increases almost one order of magnitude by decreasing the lamellae thickness in half (d). Scale bars in insets (a) and (b): 50 µm. Error bars represent the standard deviation.*

|  | **Structure feature** | **Control** | **Length scale** | **Reference of freeze-casting papers** |
|---|---|---|---|---|
| *Specific to freeze-cast materials* | Orientation and alignment of the lamella | Patterning, temperature gradient orientation | 100 microns-1 cm | This work, [21] |
| | Shape of the pores | Additives, solid content of the suspension, solvent | 5-200 microns | This work, [22,23] |
| | Thickness of lamellae, pore dimensions | Solidification interface velocity, particle size | 2-200 microns | [6,24] |
| | Surface roughness of lamellae | Additives, solidification interface velocity | 50nm-10 microns | This work |
| | Bridges between adjacent lamellae | Additives | 2-50 microns | This work |
| | Interface between ceramic and infiltrated phase | Additives, surface pre-treatment | 10-100 nm | [6,25,26] |
| | Pores interconnectivity | Solid content of the suspension (percolation of the pores) | microns | [27] |
| *Generic to ceramic* | Grain size | | 0.1-50 microns | [28] |
| | Grain boundaries | | | |

| | Reinforcement | | 0.1-0.5 microns | [28] |

*Table 1: Structural features of freeze-cast materials and their controls*